\documentclass[%
 reprint,
bibnotes,
 amsmath,amssymb,
prd]{revtex4-1}
\usepackage{lipsum}
\usepackage{geometry}        
\usepackage{graphicx}
\usepackage{amssymb}
\usepackage{epstopdf,hyperref}
\usepackage{xcolor}
\makeatletter
\def\qrr@split@result#1 #2\@qrr@split@result{\edef\erfInput{#1}\edef\erfResult{#2}}
\newcommand*{\gnuplotErf}[2][\jobname.eval]{%
    \immediate\write18{gnuplot -e "set print '#1'; print #2, erf(#2);"}%
    \everyeof{\noexpand}
    \edef\qrr@temp{\input{#1} }%
    \expandafter\qrr@split@result\qrr@temp\@qrr@split@result
}

\newcommand{\beq}{\begin{equation}}
\newcommand{\eeq}{\end{equation}}
\newcommand{\bea}{\begin{eqnarray}}
\newcommand{\eea}{\end{eqnarray}}
\newcommand{\beann}{\begin{eqnarray*}}
\newcommand{\eeann}{\end{eqnarray*}}

\begin{document}
\title{Amending the halo model to satisfy cosmological conservation laws}
\author{Alice Y. Chen}
\email{ay7chen@uwaterloo.ca}
\affiliation{Department of Physics and Astronomy, University of Waterloo, 200 University Ave W, N2L 3G1, Waterloo, Canada}
\affiliation{Waterloo Centre for Astrophysics, University of Waterloo, Waterloo, ON, N2L 3G1, Canada}
\affiliation{Perimeter Institute For Theoretical Physics, 31 Caroline St N, Waterloo, Canada}
 \author{Niayesh Afshordi}
 \email{nafshordi@pitp.ca}
\affiliation{Department of Physics and Astronomy, University of Waterloo, 200 University Ave W, N2L 3G1, Waterloo, Canada}
\affiliation{Waterloo Centre for Astrophysics, University of Waterloo, Waterloo, ON, N2L 3G1, Canada}
\affiliation{Perimeter Institute For Theoretical Physics, 31 Caroline St N, Waterloo, Canada}

\begin{abstract}
One of the most powerful tools in the arsenal of theoretical cosmologists is the halo model of large scale structure, which provides a phenomenological description of nonlinear structure in our universe. However, it is well known that there is no simple way to impose conservation laws in the halo model. This can severely impair the predictions on large scales for observables such as weak lensing or the kinematic Sunyaev-Zel'dovich effect, which should satisfy mass and momentum conservation, respectively.  For example, the standard halo model overpredicts weak lensing power spectrum by $>8\%$ on scales $>20$ degrees.  To address this problem, we present an {\it Amended Halo Model}, explicitly separating the linear perturbations from {\it compensated} halo profiles.  This is guaranteed to respect conservation laws, as well as linear theory predictions on large scales. We then provide a simple fitting function for the compensated halo profiles, and discuss the modified predictions for 1-halo and 2-halo terms, as well as other cosmological observations such as weak lensing power spectrum. Furthermore, we compare our results to previous work, and argue that the amended halo model provides a more efficient and accurate framework to capture physical effects that happen in the process of cosmological structure formation.         
\end{abstract}
\maketitle


\section{Introduction}\label{intro}

The nature and composition of dark matter has been a long-standing problem in cosmology.  All observational evidence for the existence of dark matter has so far been purely gravitational, and it is based upon these observations that we currently infer dark matter's nature and properties. Due to these observations, it is currently hypothesized that dark matter particles do not have any other detectable signatures aside from gravity.  As a result, they also do not interact with standard model particles or photons, other than through their gravitational pull.  Commonly proposed classifications for these particles include Weakly Interacting Massive Particles (WIMPs) and axions - for the model we study in this paper, either one of these particles could work as long as the particles themselves are cold, collisionless, and have negligible self-interactions.

The most commonly used analytic framework for the formation of dark matter structure in cosmology has been the Standard Halo Model (SHM), where dark matter particles clump together to form (nearly-) spherical virialized structures known as halos.  Dark matter particle properties, along with the cosmological initial conditions, determine the properties of SHM, which describes how halos are formed and what their internal structures are like (e.g., \cite{2000MNRAS.318..203S,Cooray:2002dia}). In spite of its success in describing the statistics of nonlinear structures on small scales (e.g., \cite{2018MNRAS.478.1042S}), the SHM  is not dynamical, and thus has no way to guarantee conservation laws,  such as for mass or momentum. This leads to unphysical behaviour, such as significant deviations from linear theory predictions at small wavenumbers, $k \to 0$, due to the dominance of the 1-halo term, e.g., \cite{Cooray:2002dia, 2017PhRvD..96h3528G}.  This 1-halo term mainly describes how dark matter density inside halo structures correlate with each other and hence why it is dominant on small scales but not on large scales.

\begin{figure}
\centering
    \includegraphics[width=1.1\linewidth]{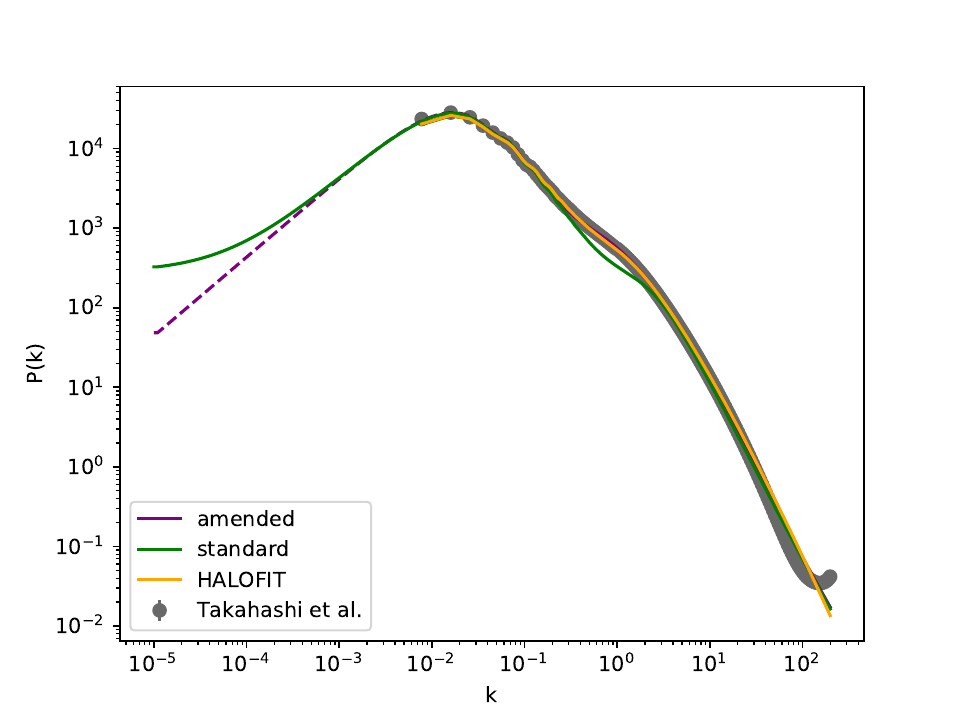}
    \caption{\footnotesize Comparison of predictions for nonlinear matter power spectrum, using Amended Halo Model (introduced here), HALOFIT \cite{Takahashi:2012em}, and the standard halo model \cite{Cooray:2002dia}. The data points are from Takahashi et al. simulations \cite{Takahashi:2012em}.  It can be seen that the standard halo model power starts to approach a constant value at small k's instead of 0, which is unphysical, as the standard model does not conserve mass. In our amended model, the power approaches 0 as k$\rightarrow$0, similar to HALOFIT and linear theory predictions.}
    \label{fig:power_extrapolate}
\end{figure}

Even though at first this may sound like an academic question, current and upcoming wide-field surveys of weak lensing, kinematic Sunyaev-Zel'dovich effect, and pre-reionization 21-cm intensity will probe total mass, momentum, and hydrogen mass on large scales.  Thus, they will be sensitive to theoretical deficiencies such as violation of conservation laws that the SHM entails.  For example, the current SHM overpredicts power in the region around the halo radius, so observed data may be misinterpreted for objects that are not dark matter.  As a result, we need to make amendments to the current halo model in order to obtain a more accurate picture of cosmic structures on all scales. This paper provides a simple and user-friendly prescription to implement this amendment, what we will call the {\it Amended Halo Model}. Our result can be visually summarized in Figure (\ref{fig:power_extrapolate}), which compares our amended halo model predictions for the matter power spectrum, to the standard halo model which overpredicts power on large scales.

The structure of the paper is as follows:  Section \ref{sec::review} briefly reviews the Standard Halo Model (SHM) and its limitations, while Section \ref{sec::AHM} outlines our amended halo model (AHM) and how it addresses issues highlighted in Section \ref{sec::review}.  Section \ref{sec::methods} describes the method we used to test AHM, Section \ref{sec::results} summarizes how well it does in fitting the cosmological simulation data, and Section \ref{sec::CMB_predictions} discusses the predictions AHM makes for weak lensing.  Section \ref{previous} briefly compares our approach to previous work. Finally, our results, implications, and future prospects for the AHM are summarized in Section \ref{sec::conclusion}.

\section{Standard Halo Model (SHM)}\label{sec::review}

The most important ingredient of the SHM is that all cosmological halos approximately follow a parametrized universal density profile.  The earliest proposal for this density profile was the Navarro, Frenk, and White (NFW) profile \cite{Navarro:1995iw}, which has been around since the 1990s, although more precise extensions have been considered more recently \cite{Graham:2005xx,Tavio:2008em, Hadzhiyska:2019xnf}.
The NFW profile is the most commonly used profile to date, and was developed through N-body simulations of dark matter particles  by Navarro, Frenk, and White (NFW) \cite{Navarro:1995iw}.  By using the data from these large scale simulations, they came up with a formula that describes the spherically averaged density of dark matter within halo structures.  This density was fitted by:

\beq
\delta_{\rm NFW}(r) \equiv \cfrac{\rho (r)}{\bar{\rho}} = \cfrac{\Omega_m\delta_c}{(r/r_s)(1 + r/r_s)^2},\label{delta_nfw}
\eeq
where $\rho(r)$ is the density of the halo region, $r$ is the radius from halo centre, and $\bar{\rho}= \frac{3\Omega_m H^2}{8\pi G}$ is the mean density of the universe.  ${\delta}_c$ is defined as

\beq
\delta_c = \frac{200c^3}{3\left[\ln(1+c) - c/(1+c)\right]},
\eeq
 
 where $c=r_{\rm 200c}/r_s$ is the concentration parameter, defined as the ratio of $r_{\rm 200c}$ (within which, mean halo density  $= 200\times \rho_{\rm crit}$),   and $r_s$ (known as the scale radius).

The NFW profile provides a good model for dark matter density inside the virialized halo region; however, significantly outside this region (i.e. on large scales), where $r$ $\gg$ $r_{\rm vir}$, there does not exist a clear consensus on a universal dark matter density profile (see \cite{Tavio:2008em} for one proposal).  This is one of the main reasons why we need large scale dark matter distribution models, such as SHM or AHM, beyond the NFW profile - the NFW profile is limited in what it can describe.  While NFW works well inside halos in dynamical equilibrium, it is not a good fit for dark matter density beyond the halo radius or for structures not in dynamical equilibrium.

 In SHM, the matter overdensity is written as a sum over halos:
\beq
\delta({\bf x}) = \sum_j \delta^j({\bf x - x}_j), \label{delta_r}
\eeq
or in Fourier space as:
\beq
\delta_{\bf k} = \sum_j \delta^j_{\bf k} \exp(i{\bf k\cdot x}_j).
\label{delta_f}
\eeq

For individual halo profiles in SHM, we often use the Fourier transform a mean profile (assuming its universality): 
\beq
\delta^j_{\bf k} \simeq \frac{M^j}{\bar{\rho}} u(k|M^j) \equiv  \int d^3{\bf x} \exp(i{\bf k\cdot x}) \delta^j_{\rm mean}(|{\bf x}| | M^j). \label{u_av}
\eeq
$M^j$ here is not the total mass of the $j$-th halo, which is not well-defined to begin with, but rather the mass on a fixed scale. We opt to use $M_{200c}$ as the mass within the radius where the mean halo density is 200$\times \bar{\rho}/\Omega_m$. 

Note that  Equation (\ref{u_av}) ignores the (possibly correlated) variations in profiles of halos with the same $M^j$, which is a fundamental limitation of the SHM, and halo models in general. We shall come back to this issue, and our quick fix for it, below.  

For NFW profile (\ref{delta_nfw}), $u(k|M)$ has the analytical form: 
\begin{widetext}
\beq
u_{\rm NFW}(k|M) = \frac{4\pi {\rho}_s r_s^3}{M}\left\{ \sin(kr_s) \big({\rm Si}[(1+c)kr_s] - {\rm Si}(kr_s)\big) + \cos(kr_s)\big( {\rm Ci}[(1+c)kr_s] - {\rm Ci}(kr_s)\big) - \frac{\sin(ckr_s)}{(1+c)kr_s} \right\},
\label{shm_density}
\eeq
\end{widetext}
where Si and Ci are the sine and cosine integral functions respectively \cite{Cooray:2002dia}.  We find the concentration parameter c using the Equations (56)-(57) of Okoli \& Afshordi's 2015  \cite{Okoli:2015dta}.  We then use this $c$ to then find the scale radius from NFW.

Let us now discuss the simplest application of the halo model. Given a choice of halo profile $u(k|M)$, the matter power spectrum in SHM is given by \cite{Cooray:2002dia}:
\begin{widetext}
\beq
P_{\rm SHM}(k) = \frac{1}{\bar{\rho}^2}\int dM n(M) M^2 |u(k|M)|^2 + \left[\frac{1}{\bar{\rho}}\int dM M n(M) b(M) u(k|M) \right]^2 P_L(k),
\label{shm_power}
\eeq
\end{widetext}
where the first (second) term is known as the 1-halo (2-halo) term, corresponding to density correlation within (in-between) halos. Moreover, $n(M)$ is the halo mass function \cite{2010ApJ...724..878T}, $b(M)$ is the bias function \cite{Tinker:2008ff, 2010ApJ...724..878T}, and $P_L(k)$ is the linear matter power spectrum.

As we discussed in Section \ref{intro},  there is no simple mechanism in SHM to enforce conservation laws on large quasilinear scales. It is arbitrary to split the density into multiple halos for small k's, or on large distances that involve several halos.  However, requiring $\delta_{\bf k} \rightarrow \delta_{L,{\bf k}}$ for small k's (i.e. approximately linear evolution on large scales) will also require a fine-tuned cancellation between the diagonal and off-diagonal parts of the covariance matrix of $\delta^j_{\bf k}$, for 1-halo and 2-halo terms.  For example, this would not be satisfied by the choice of a universal profile, such as NFW (\ref{u_av}), because NFW does not model dark matter density well beyond the halo virial radius.

While such a constraint is hard to impose in SHM (but see Section \ref{previous} for a summary of other attempts), in the next section, we develop an {\it Amended Halo Model} that automatically satisfies this constraint as $k \to 0$, and yet replicates the success of SHM at large k's. 

\section{Amending the Halo Model}\label{sec::AHM}

Here, we propose a small improvement to the halo model that automatically satisfies mass conservation.  As momentum is dependent on mass, this change will help conserve momentum as well.

To do this, we change Equations (\ref{delta_r}-\ref{delta_f}) to separate the linear overdensity from (now compensated) halo profiles: 
\beq
\delta({\bf x}) = \delta_L({\bf x})+ \sum_j \delta^j({\bf x - x}_j)
\label{delta_x}
\eeq
and thus
\beq
\delta_{\bf k} = \delta_{L,{\bf k}}+ \sum_j \delta^j_{\bf k} \exp(i{\bf k\cdot x}_j).
\label{delta_k}
\eeq
We also modify halo profiles u($k|m$) to become 
\beq
u_{\rm AHM}(k|M) \rightarrow f(kr_s)\tilde{u}_{\rm NFW}(k|M),
\label{u_extrapolate}
\eeq
where $f(x)$ is a dimensionless fitting function we find using simulation data. Now, requiring that $f(x) \to 0$ as $x \to 0$ ensures that individual halo profiles are {\it compensated}, i.e. have zero integral:

\beq
\int d^3{\bf x}~ \delta^j({\bf x}) = \lim_{{\bf k} \to 0}\delta^j_{{\bf k}} = \frac{M^j}{\bar{\rho}} u_{\rm AHM}(0|M^j) = 0.
\eeq

Furthermore, $\tilde{u}_{\rm NFW}$ is defined to be the same as $u_{\rm NFW}$ for large k's, but without the sharp cutoff at $r_{200c}$. In other words, we replace the sharp real-space cutoff at virial radius in AHM, with a gentle Fourier space cutoff $f(x)$, that smoothly interpolates between overdense and underdense regions. As such, we let c $\rightarrow$ $\infty$ (and thus $r_{200c}$ $\rightarrow$ $\infty$) only within the curly brackets in Equation (\ref{shm_density}) (not changing the prefactor):

\begin{widetext}
\beq
\Tilde{u}_{\rm NFW}(k|M) \simeq \frac{4\pi {\rho}_s r_s^3}{M}\left\{\sin(kr_s) \big(\frac{\pi}{2} - {\rm Si}(kr_s) \big) - \cos(kr_s){\rm Ci}(kr_s)\right\},
\label{ahm_density}
\eeq
\end{widetext}

Now, the power spectrum becomes
\begin{widetext}
\beq
P_{\rm AHM}(k) =  \frac{1}{\bar{\rho}^2}\int dM n(M) M^2 |u_{\rm AHM}(k|M)|^2+ \left[1+ \frac{1}{\bar{\rho}}\int dM M n(M) b(M) u_{\rm AHM}(k|M) \right]^2 P_L(k),
\label{ahm_power}
\eeq
\end{widetext}

This new power spectrum will automatically approach linear power when $k \to 0$, as $\tilde{u}_{\rm AHM}(k|M) \to 0$, but will recover SHM on large k's with small corrections.  In the next section, we find that this amended model gives a better fit at small k's than the standard halo model does, based on data from N-body simulations.  It also yields fits on the same level of accuracy as the numerical HALOFIT package \cite{Peacock:2000qk, Takahashi:2012em}, based on a more solid physical picture of structure formation.

\section{Method and Simulations} \label{sec::methods}

In order to compare with the HALOFIT model used in CAMB  package \footnote{\url{https://camb.info/}}\cite{Lewis:1999bs}, the data used to investigate this amended halo model was obtained through N-body simulations of dark matter evolution, using Gaussian $\Lambda$CDM linear initial conditions.  The simulation data is primarily from Takahashi et al.'s paper (\cite{Takahashi:2012em}, using Nishimichi's simulations). We studied the matter power spectra at $z=0$ for different cosmologies, summarized in Table \ref{tab: cosmology}.
\begin{table}
 \begin{tabular}{||c | c | c | c | c | c||} 
 \hline
 \hspace{3mm} & $\Omega_{b}$ & $\Omega_{m}$ & h  & $\sigma_{8}$ & $n_{s}$  \\ [1.0ex] 
 \hline\hline
 WMAP1 & 0.044 & 0.290 & 0.72 & 0.90 & 0.99 \\ 
 \hline
 WMAP3 & 0.041 & 0.238 & 0.732 & 0.76 & 0.958 \\
 \hline
 WMAP5 & 0.046 & 0.279 & 0.701 & 0.817 & 0.96 \\
 \hline
 WMAP7 & 0.046 & 0.290 & 0.70 & 0.81 & 0.97 \\ [1.0ex] 
 \hline
\end{tabular}
\caption{\footnotesize Cosmology parameters used in Takahashi et al.'s simulations \cite{Takahashi:2012em}. }
\label{tab: cosmology}
\end{table}
The simulations had box sizes of 320, 800, and 2000 Mpc/h and particle number of $1024^{3}$, starting at redshift $z=99$ and ending at $z=0$.

\section{Results and Discussion}\label{sec::results}

We apply Equations (\ref{delta_x}-\ref{ahm_power}) to the power spectra obtained from the simulation data of Takahashi et al.'s \cite{Takahashi:2012em} (used to calibrate the HALOFIT model) and attempt to parametrize the cutoff function $f(x)$ in (\ref{u_extrapolate}) that can fit the data with an error $\leq$ 5$\%$. 

Furthermore, we require $f(x) \propto x^2$ for $x\ll 1$, while it approaches $1$ for $x \gg 1$. The former ensures that the (spherically) averaged halo profile is analytic in ${\bf k}$ and compensated, i.e. the leading term in $\delta_{\bf k}$ should be ${\bf k \cdot k}$ in the Taylor expansion. The latter ensures that we recover SHM with NFW profiles on small scales/large $k$'s.  

We find that the following parametrization for $f(x)$ satisfies these requirements:
\beq
f(x) = \frac{ax^2 + bx^3 + dx^4}{1+ cx^3 + dx^4 },
\label{f_fit}
\eeq
where the best-fit parameter values for $a$, $b$, $c$, and $d$ are listed in Table \ref{tab: params}.  The fits are found by minimizing the root-mean-squared of relative errors, defined as:
\beq
{\rm Error} \equiv 2\sqrt{\left\langle \left[ P(k)^{\rm sim}_{i} - P(k)^{\rm model}_{i} \over P(k)^{\rm sim}_{i} \right]^2 \right\rangle_i} ,
\eeq
where the average is over the simulated data points in $k$-space. Meanwhile,  $P(k)^{\rm sim}_{i}$ is the power spectrum from simulation data, and $P(k)^{\rm model}_{i}$ is the theoretical power spectrum from either (\ref{shm_power}) or (\ref{ahm_density}), depending on whether we are finding the error for the standard halo model (SHM) or amended halo model (AHM).

The first parameter $a$ also has the physical significance of being related to the second moment of the compensated halo profile, i.e. Taylor expanding Equation (\ref{u_av}) in ${\bf k}$, we can see that:
\beq
a = -\frac{\bar{\rho}}{6M^j r^2_s} \int d^3{\bf x}|{\bf x}|^2 \delta^j_{\rm AHM}(|{\bf x}| |M).
\eeq

Since the compensated halo profile is overdense in the middle, and underdense in the outskirts, we expect the 2nd moment to be negative, and thus $a>0,$ as seen in our best fits in Table \ref{tab: params}.

\begin{table}
    \centering
    \begin{tabular}{||c | c | c | c | c | c||} 
    \hline
      & $a$ & $b$ & $c$ & $d$ \\ 
    \hline\hline
    WMAP1 & 0.018 & 10.5 & 10.7 & 3.03 \\
    \hline
    WMAP3 & 1.94 & 20.2 & 21.6 & 0.0034 \\
    \hline
    WMAP5  & 0.453 & 18.5 & 19.0 & 0.0055 \\
    \hline
    WMAP7 & 0.577 & 18.4 & 18.9 & 0.0286 \\
    \hline
    Mean & 0.747 & 16.9 & 17.55 & 0.767 \\
    \hline
    Stn Dev. & 0.72 & 3.76 & 14.1 & 1.3 \\
    [1.5ex] 
    \hline
    \end{tabular}
    \caption{\footnotesize  Table: Fitting parameters for (\ref{f_fit}) for the different WMAP cosmologies from \cite{Takahashi:2012em}, with the average and standard deviation for each parameter listed in the last two rows. }
    \label{tab: params}
\end{table}
 
\begin{figure*}
    \begin{tabular}{|c|c|}
    \hline
    WMAP1 & WMAP3  \\
    \includegraphics[width=.45\linewidth]{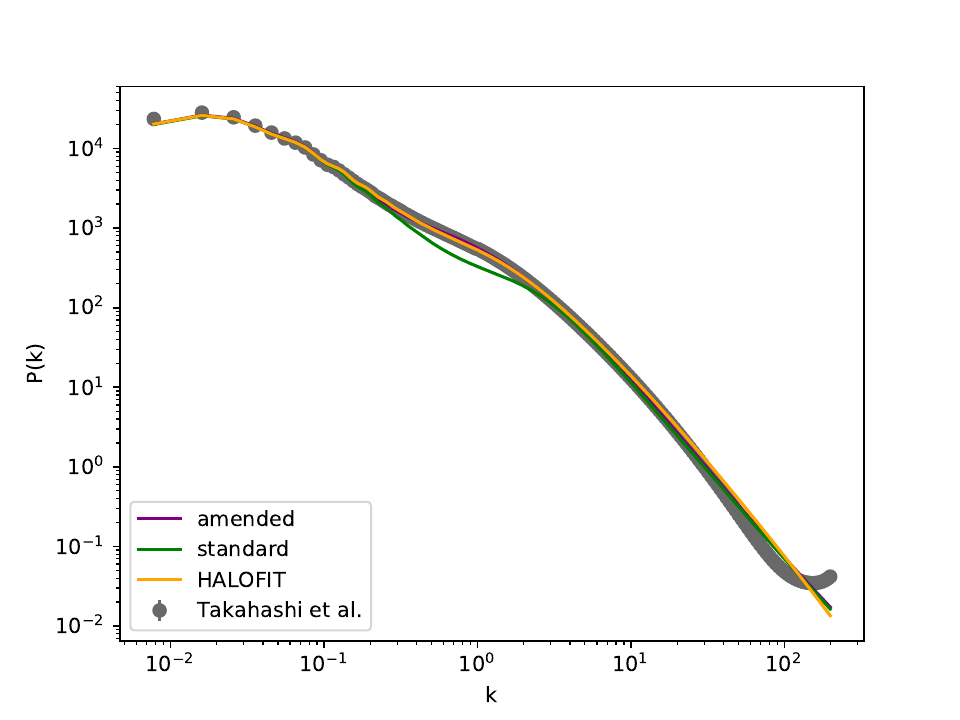} & \includegraphics[width=.45\linewidth]{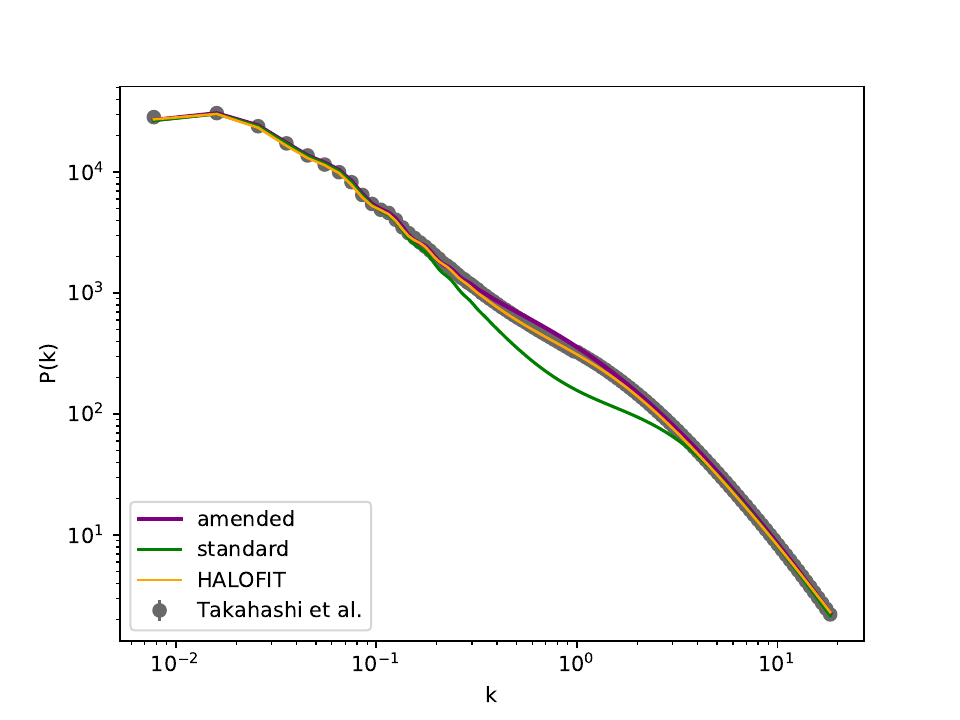}  \\
    \includegraphics[width=.4\linewidth]{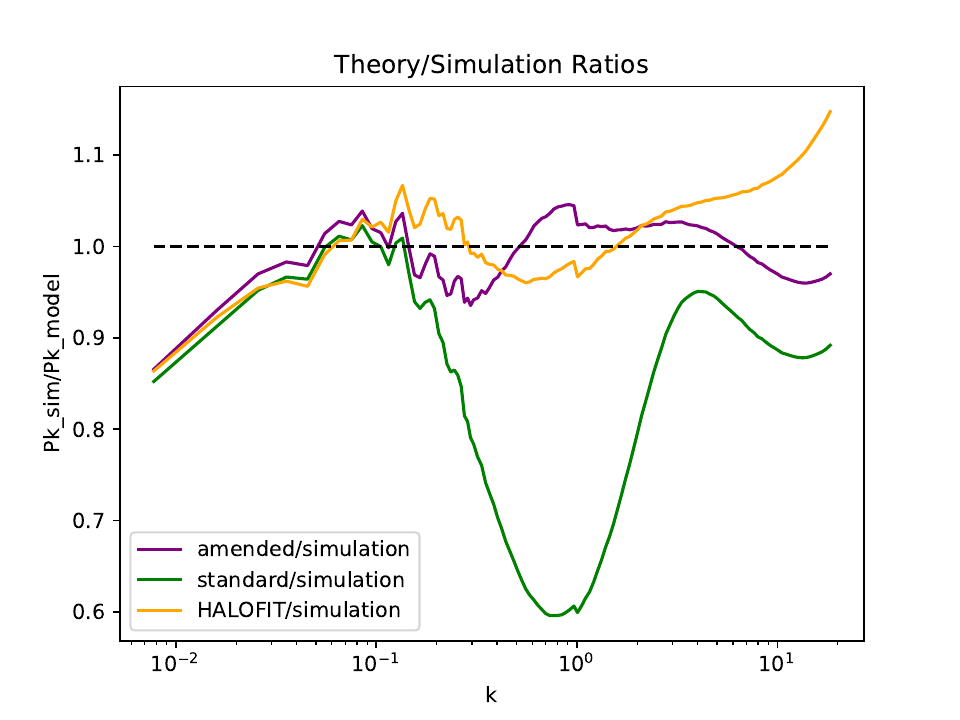} & \includegraphics[width=.4\linewidth]{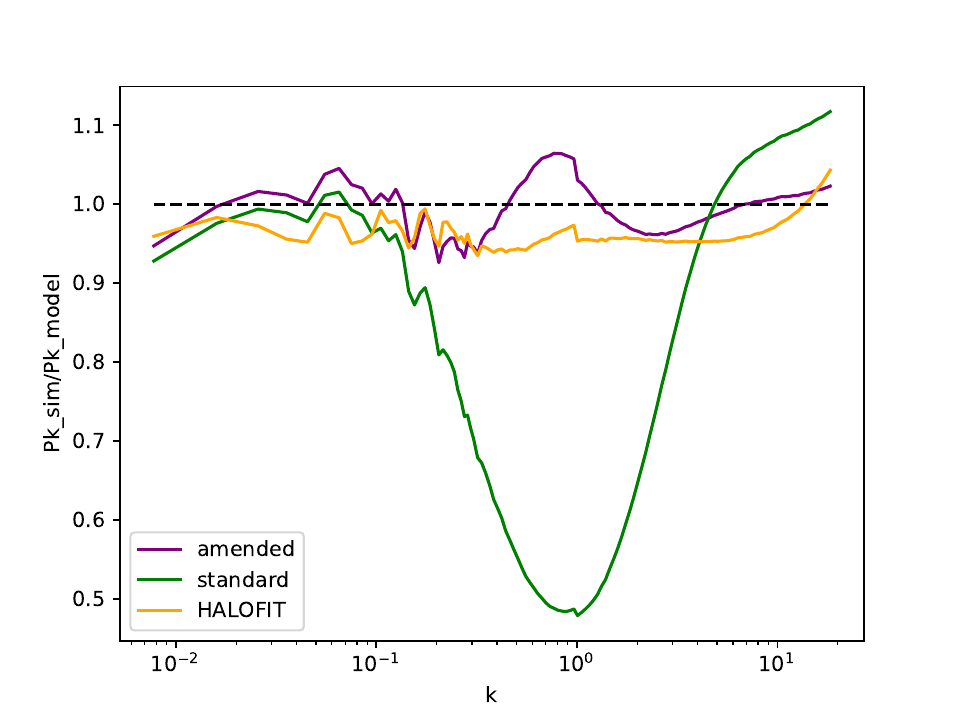}  \\
    \hline
    WMAP5 & WMAP7  \\
    \includegraphics[width=.45\linewidth]{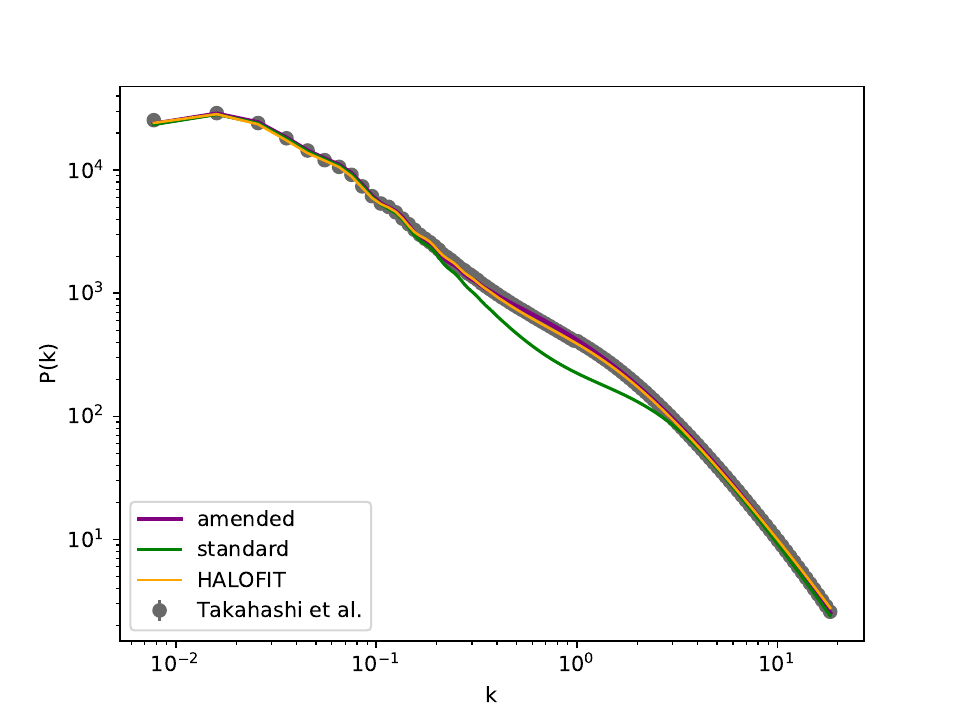} & \includegraphics[width=.45\linewidth]{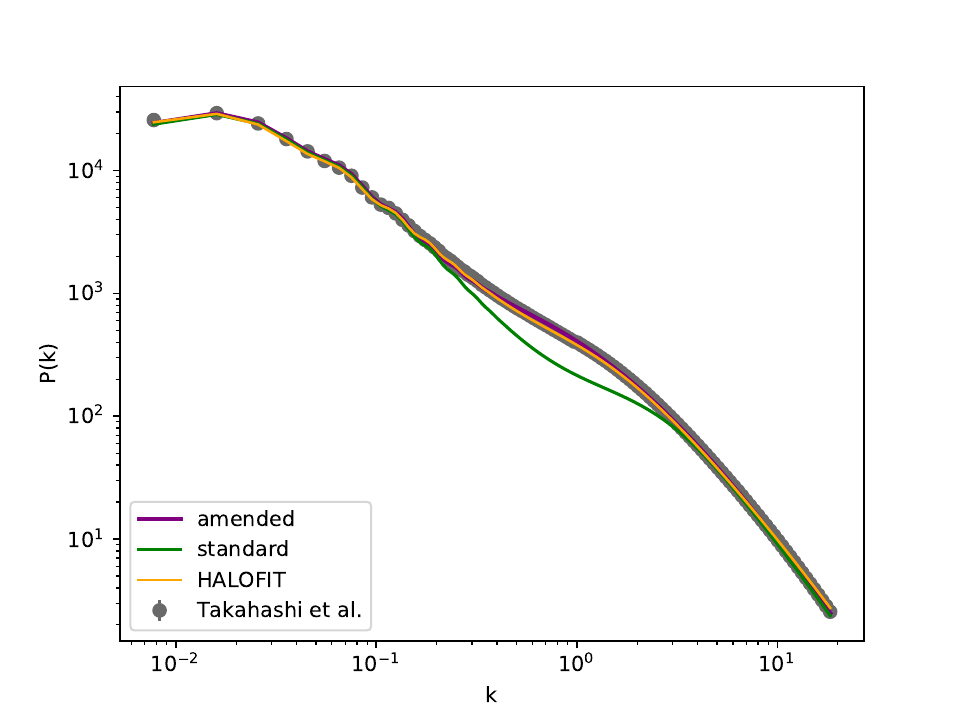}  \\
    \includegraphics[width=.4\linewidth]{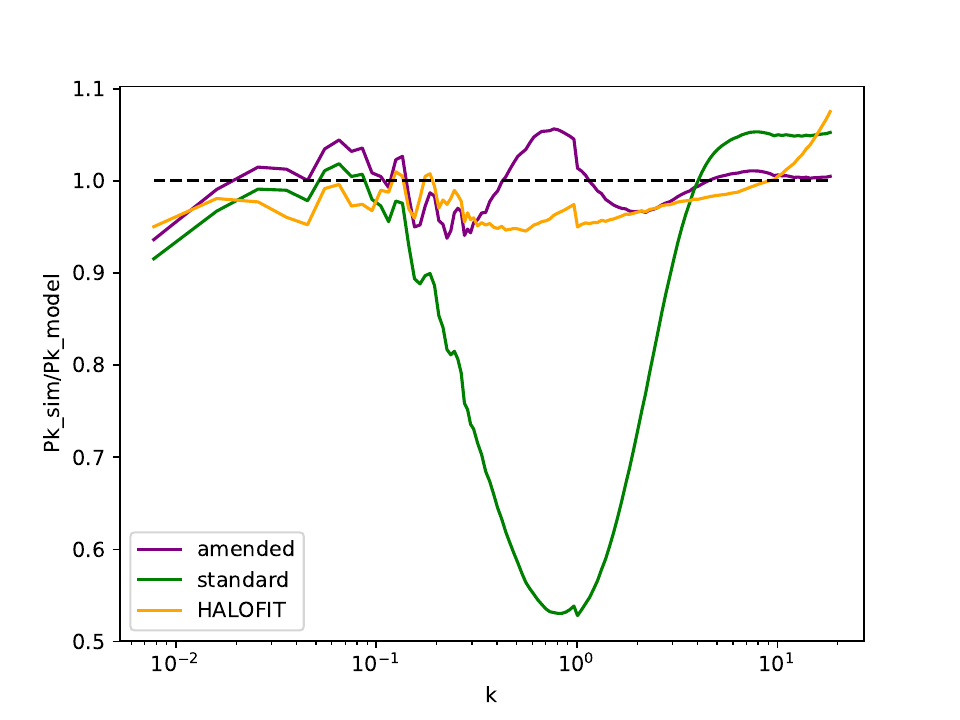} & \includegraphics[width=.4\linewidth]{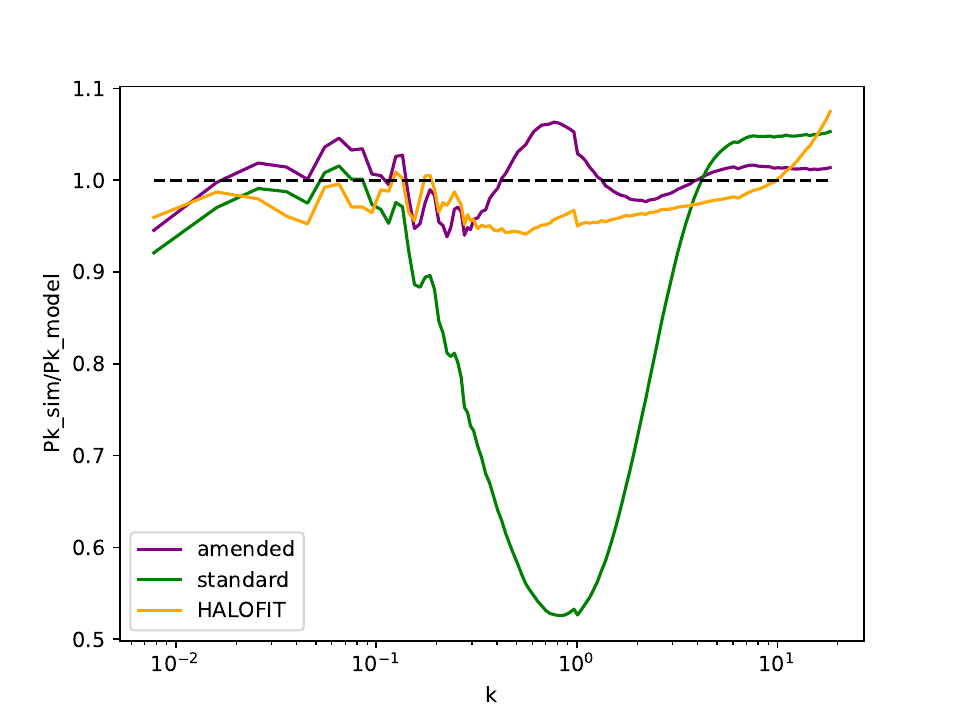}  \\
    \hline
    \end{tabular}
    \caption{\footnotesize  Comparison of Takahashi et. al's simulation data  \cite{Takahashi:2012em} with the standard halo model (SHM), amended halo model (AHM), and HALOFIT.  The panels below each $P(k)$ plot show the ratios of model to simulated spectra.}
\label{fig:test}
\end{figure*}

The resulting mean relative errors for different simulations are summarized in Figure \ref{fig:test}. Table \ref{tab: some errors} compares the relative errors with those of HALOFIT, assuming that we use the best-fit parameters from Table \ref{tab: params} for each simulation. We see that, while we achieve smaller errors compared to HALOFIT, we also have more parameters per simulation (4/sim for this work, versus 35/ 16 sim's in \cite{Takahashi:2012em}). If we fix all parameters to their average over 4 simulations, effectively having 1 parameter per simulation, Tabel \ref{tab: avg errors} shows that we get generally larger errors than HALOFIT. Therefore, as a fitting function, AHM using Equation (\ref{f_fit}) has a comparable performance to HALOFIT, while it is based on a more physical underlying framework. We also see that both AHM and HALOFIT do far better than the SHM in fitting the simulated data.  

We can also use Akaike Information Criterion (AIC) to compare HALOFIT and AHM.  The AIC is given as
\beq
AIC = 2\kappa - 2\ln(\hat{L})
\eeq 
where $\kappa$ is the number of parameters a model uses, and $\hat{L}=\exp(-\chi^2_{\rm min}/2)$ is its maximum likelihood.  The average AIC for our amended model for the individually fitted parameters is around 17,000, while for the mean parameters the amended AIC is around 70,000.  The average AIC for Takahashi et al.'s model is around 19,300 - a lot better than AHM using mean parameters but slightly worse than AHM's individually fitted parameters.  However, this comparison should be taken with a grain of salt as we are using the parameters found to minimize relative error, not the $\chi^2$, to compute AIC.

\begin{table}
    \centering
    \begin{tabular}{||c | c | c | c | c ||} 
    \hline
    \hspace{3mm} & Standard & HALOFIT & Amended \\ [1.0ex] 
    \hline\hline
    WMAP1  & 0.23 & 0.053 & 0.027 \\
    \hline
    WMAP3  & 0.29 & 0.042 & 0.033 \\
    \hline
    WMAP5  & 0.25 & 0.034 & 0.029 \\
    \hline
    WMAP7  & 0.25 & 0.037 & 0.029 \\ 
    [1.5ex] 
    \hline
    \end{tabular}
    \caption{\footnotesize  Mean relative errors for the Standard Halo Model (SHM), HALOFIT, and Amended Halo Model (AHM) for the different WMAP cosmologies in Figure (\ref{fig:test}), using each cosmology's individually optimized parameters (Table \ref{tab: cosmology}). }
    \label{tab: some errors}
\end{table}

\begin{table}
    \centering
    \begin{tabular}{||c | c | c | c||} 
    \hline
    \hspace{3mm} & Standard & HALOFIT & Amended  \\ [1.0ex] 
    \hline\hline
    WMAP1  & 0.23 & 0.053 & 0.11 \\
    \hline
    WMAP3  & 0.29 & 0.042 & 0.08 \\
    \hline
    WMAP5  & 0.25 & 0.034 & 0.031 \\
    \hline
    WMAP7  & 0.25 & 0.037 & 0.029 \\ 
    [1.5ex] 
    \hline
    \end{tabular}
    \caption{\footnotesize  Errors for the Standard Halo Model (SHM), HALOFIT, and Amended Halo Model (AHM) if we use the average parameters for all the different WMAP cosmologies, instead of their individually optimized ones. }
    \label{tab: avg errors}
\end{table}

\begin{figure}
\centering
    \includegraphics[width=0.95\linewidth]{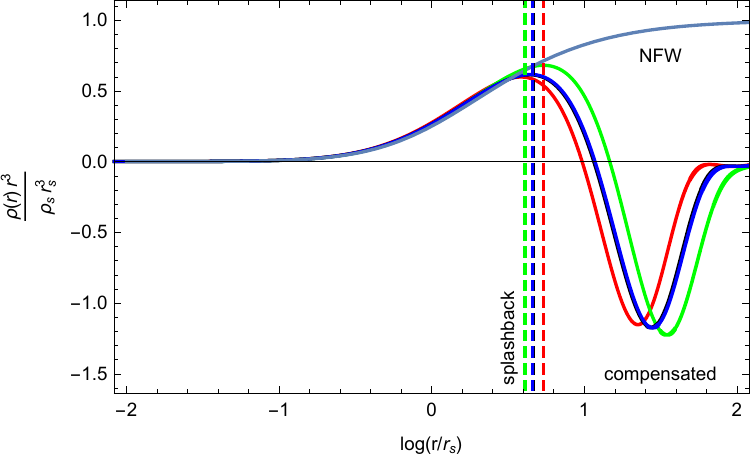}
    \caption{\footnotesize The best-fit compensated density profiles for the amended model, in different cosmologies, compared to the NFW profile.  The light blue curve at the top represents the NFW density profile, while the red, green, black, and blue curves are the densities obtained from our amended model for the WMAP1, WMAP3, WMAP5, and WMAP7 cosmologies respectively.  The vertical lines are for the splashback radius from More, Diemer, and Kratsov  \cite{Diemer:2014xya, More:2015ufa}, which is close to where density starts decreasing rapidly in our model as well. }
    \label{fig:NFW_FFT}
\end{figure}

To get a more physical picture, we can look at the dark matter density that we obtain from (\ref{ahm_density}) by using an inverse Fourier Transform.  On smaller scales, inside the halos (at distances smaller than the halo's $r_{200c}$), we should roughly recover the NFW density profile.  However, outside the virial radius of a halo, we should expect the amended "compensated" profile density to go negative in order to satisfy mass conservation.  From Figure \ref{fig:NFW_FFT}, we see that our density profile does match NFW up to the mean $r_{200c} \sim 6 \times r_s$, which is what we would expect from theory. However, it crosses zero and becomes negative at roughly roughly $2\times r_{200c}$, although the exact value appears to depend on cosmology.

From Figure \ref{fig:test}, it can be seen that the deviation from simulation data (from \cite{Takahashi:2012em}) resulting from our amended model is significantly smaller than the deviation from the standard halo model, indicating that this new modified power spectrum is a better fit for dark matter density in general.  When k is a large enough number - k $\geq$ 5 h/Mpc - the amended model, the standard halo model, and the numerical HALOFIT all produce similar results, as we should expect given that HALOFIT and the amended model are supposed to replicate SHM on large k's.  However, as seen in Figure \ref{fig:test}, AHM and HALOFIT are significantly more accurate than SHM on intermediate, scales with $k \sim 1$ h/Mpc.  Physically, this indicates that the current halo model profile does require some compensation to fit the data (!), similar to what we proposed in (\ref{delta_x})-(\ref{delta_k}).  This new halo model also conserves mass and fits the simulated dark matter density as well as previous models, resulting in a new physical model for dark matter clustering on large scales. 

Another advantage of AHM is that, unlike SHM, it has little sensitivity to including small halos. The reason is that in SHM, it is assumed that all the mass is included in halos, and therefore convergence of integrals over halo mass requires including relatively small halos. However, in AHM the halos are compensated (i.e. have zero mass), and thus small halos do not contribute to large scale observables. 

\section{CMB Lensing}\label{sec::CMB_predictions}

We provide an example of how mass non-conservation can impact observational predictions.  In this section we shall study the weak lensing of cosmic microwave background (CMB) maps, that is being measured with unprecedented precision using current and future experiments (e.g., \cite{Sherwin:2016tyf, Omori:2017tae,Aghanim:2018oex}). To see what power AHM would predict, and show that SHM overpredicts lensing power, we calculated the weak lensing power that should be observed from the Cosmic Microwave Background (CMB) \cite{Aghanim:2018oex, Ade:2015zua} using AHM and the Extended Limber approximation \cite{LoVerde:2008re}.

\begin{figure}
\centering
\includegraphics[width=0.5\textwidth]{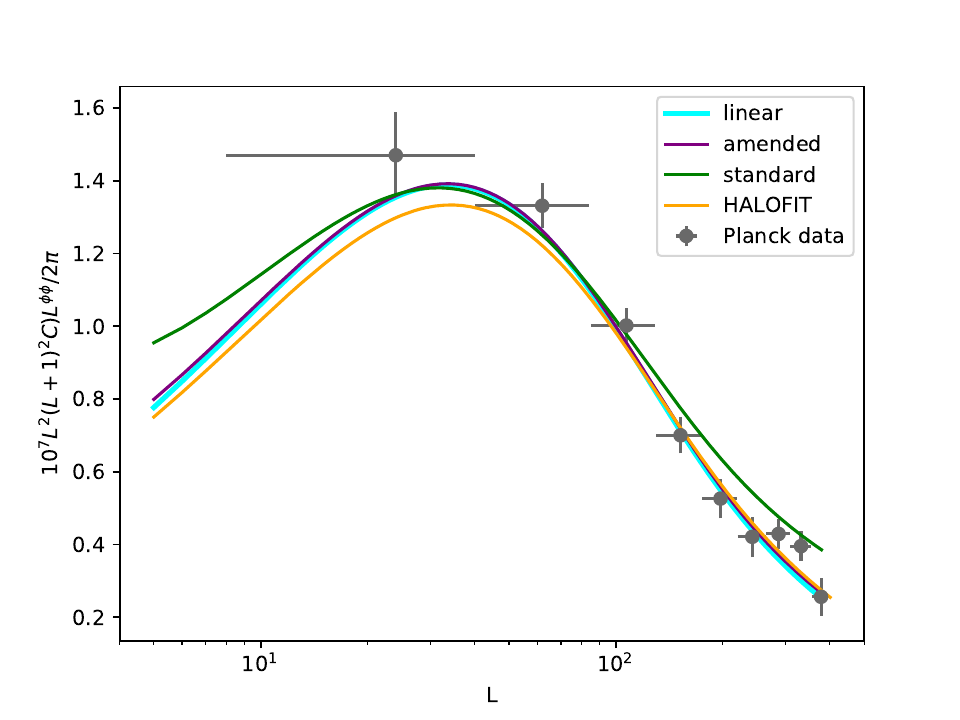}
\caption{\footnotesize  Predictions for CMB gravitational lensing power spectra for our amended halo model (AHM), standard halo model (SHM), and linear power (while the linear power spectrum is obtained from HALOFIT, this is the initial power spectrum from a high redshift, and not the same one we graphed in Figure 2 of present day).  The cosmology used here is WMAP7 since it has the closest parameters to the 2018 Planck cosmological parameters \cite{Aghanim:2018eyx}.  For comparison, we show the measurement of CMB power spectrum from Planck 2018 (plotted as the grey errorbars) \cite{Aghanim:2018oex}.  It can be seen that the SHM generally overpredicts power compared HALOFIT and AHM, on small and large L's. This is because of non-conservation of mass on large scales in the SHM.}
\label{fig:cmb_lensing}
\end{figure}

\begin{figure}
\centering
    \includegraphics[width=1.1\linewidth]{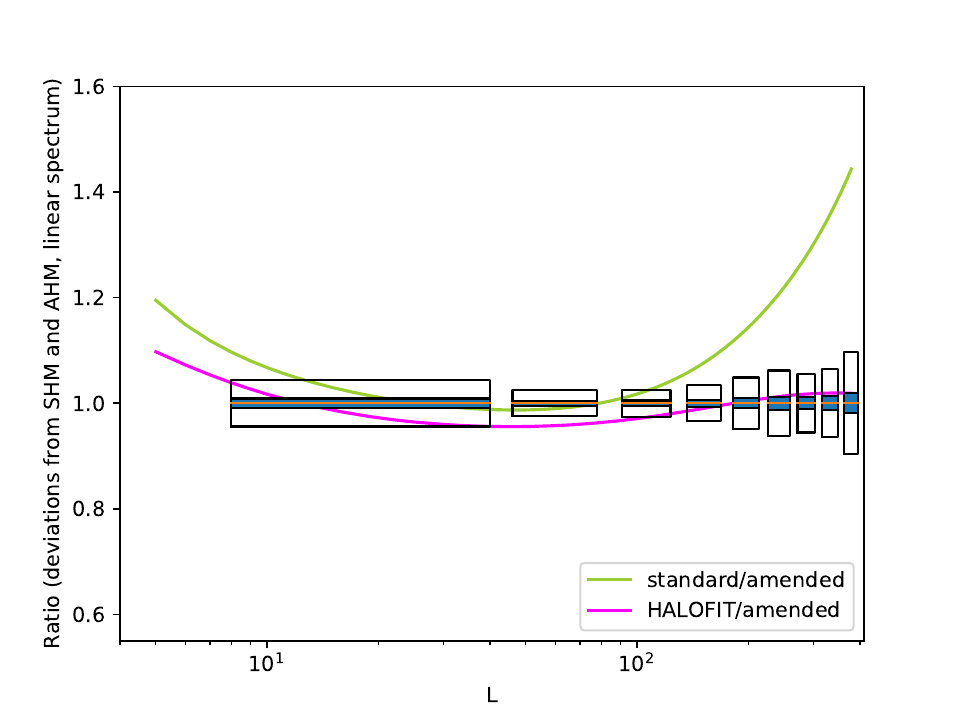}
    \caption{\footnotesize Ratios of the predicted lensing powers SHM and HALOFIT to our proposed amended model (AHM).  The white boxes show the Planck errors, while the blue boxes show CMB-S4 projected errors \cite{Aghanim:2018oex, Abazajian:2016yjj}. We see that SHM is already disfavored at $L \gtrsim 100$ by Planck, while its discrepancy at $L \lesssim 20$ will be probed by CMB-S4. }
    \label{fig:cmb_ratio}
\end{figure}

This lensing power, as can be seen from Figure \ref{fig:cmb_lensing}, matches the measurements from 2018 Planck results \cite{Aghanim:2018oex} fairly well.  The standard model does seem to overpredict the power (Fig. \ref{fig:cmb_ratio}), as a result of the large 1-halo contribution to power spectrum at high redshifts. We believe this is primarily because of a lack of mechanism for mass conservation in the SHM.

\section{Comparisons to Previous Alternative Models}\label{previous}

In this section, we briefly summarize two other approaches to extending the halo model, which share some of the properties of the AHM. 

In \cite{Schmidt:2015gwz}, Schmidt introduces an Effective Field Theory approach to the halo model, or ``{\it EHM}'', that includes stochastic halo additive and multiplicative terms.  As we discussed in Section \ref{sec::review}, modelling the covariance of this stochasticity is then the key ingredient in ensuring mass and momentum conservation. While EHM might be more realistic and general than the simple version of the AHM presented here, it requires modelling more free functions.  EHM also does not model the transition region where the 1-halo term domination in the power spectrum ends and the 2-halo term starts to take over, which is addressed in AHM in this paper.  From Figure \ref{fig:test}, it can be seen that even during the transition region where SHM (0.5 h/Mpc $\lesssim k \lesssim$ 2.0 h/Mpc) becomes inaccurate, AHM still fits the data well.

Another similar approach was presented in \cite{Seljak:2015rea,Hand:2017ilm}, where the matter power spectrum is modelled as the sum of the power evaluated using the Zel'dovich approximation, plus an effective ``compensated'' 1-halo term. While this provides a good fit to simulations for $k \lesssim 1$ h/Mpc, with a relatively small number of parameters, it is not clear that it can be interpreted as a consistent halo model, as the compensated halo profile is not included in the Zel'dovich spectrum (which replaces the 2-halo term). For example, finding the bispectrum would require introducing additional free functions. One bi-product of this inconsistency might be the large halo compensation scale of $\sim 26$ Mpc/h, which is significantly larger than the sizes of halos in Lagrangian coordinates. In contrast, our compensation scale (e.g., the minimum of $\rho(r) r^3$ in Figure \ref{fig:NFW_FFT}) is $\sim 5$ Mpc/h for $10^{14} M_{\odot}$/h halos, which is comparable to the Lagrangian radius of $\sim 7$ Mpc/h for these halos. Furthermore, the approaches of \cite{Seljak:2015rea,Hand:2017ilm} are built as Pade expansions in $k$, and unlike AHM (or SHM), cannot be extended to the deeply nonlinear regime.

\section{Conclusions and Future Prospects}\label{sec::conclusion}

In this paper, we introduced the amended halo model (AHM) of structure formation, which fixes the problem of mass non-conservation in the standard halo model (SHM), utilizing a simple and user-friendly framework. 
The compensated halo profiles in AHM provide predictions for the matter power spectrum (\ref{ahm_power}) that fit N-body simulation data as well as the parametrized HALOFIT model used in CAMB, and better than the standard halo model for mid to small values of k (k $\leq$ 5 h/Mpc).  This leads to an accurate and physical halo model that conserves mass and fits simulations and theoretical expectations, on both small and large scales. 

Momentum conservation here is expected mostly a result of mass conservation.  Since momentum is dependent on mass, conserving mass also results in conserving momentum, if we assign a uniform velocity to each halo.  On large scales, momentum distribution is the same in linear theory, but halos are expected to have their own compensated momentum profiles, a topic that we shall defer to future study (see below).

While the current model introduces the AHM, several future steps can be anticipated:
\begin{enumerate}
    \item Future work is needed to see whether AHM can be further developed to include halo substructure.
    \item Since our halos are compensated, the lowest order multipole moment is {\it dipole}. While the mean dipole would vanish for an average profile, it could have a scatter that contributes to the matter power spectrum on large quasilinear scales.  It would be interesting to hunt for this dipole signal in simulations or weak lensing observations.  
    \item More generally, should (co-)variance of halo profiles be included in the AHM framework, and if so, how? 
    \item Can we use match-filter methods to directly measure mean compensated halo profiles from N-body simulations?
    \item The AHM framework can be further fine-tuned and/or tested using larger boxes, as well as neutrino and/or baryonic effects. 
    \item Similar to CMB lensing studied here, it would be interesting to see how predictions for the kinematic Sunyaev-Zel'dovich effect and 21cm intensity mapping at high-z, which may be sensitive to momentum and hydrogen mass conservation, might be impacted.
    \item Another potential application of AHM is capturing environmental dependence of halo properties through cross-terms such as$\colon$
    \beq
     \sum_j\langle \delta_L({\bf x}') \delta^j({\bf x}-{\bf x}^j) \rangle,
    \eeq
    that contribute to 2-point correlation function (or the power spectrum). This could be further generalized to other tracers, such as galaxies or hot gas, by quantifying how profiles of individual halos may be different for environments with different linear overdensities.
    \item AHM can be used to model 1-point probability distribution function for conserved observables, such as weak lensing convergence, or kinetic Sunayev-Zel'dovich effect (e.g., extending treatment introduced in \cite{Thiele:2018jdl}).  
    \item Momentum conservation here is a result of mass conservation, but halos can have their own compensated momentum profiles, which is not addressed here.  Exploring how this momentum profile will fit into AHM can give further physical evidence and insights for the need of halo model amendments.
\end{enumerate}

\section*{Acknowledgments} \label{sec::acknowledgments}
We would like to thank Perimeter Institute and the University of Waterloo for their support in this work.  In particular, we would like to thank Erik Schnetter, Ethan Campbell, and Dustin Lang for their help in the use of the Symmetry supercluster at Perimeter to run simulations.

We would also like to thank Ryuichi Takahashi, Masanori Sato, Takahiro Nishimichi, Atsushi Taruya, and Masamune Oguri for letting us use their large scale simulation data in this model.

Finally, we would like to thank Neal Dalal, Vincent Desjacques, Fabian Schmidt, Uros Seljak, Kendrick Smith, and Leander Thiele for helpful discussions and comments.

For calculating linear and numerical nonlinear power spectra for comparion with our model, we used the CAMB package in Python \cite{Lewis:1999bs, Peacock:2000qk, Takahashi:2012em}.

\bibliography{HaloRef}

\begin{thebibliography}{28}%
\makeatletter
\providecommand \@ifxundefined [1]{%
 \@ifx{#1\undefined}
}%
\providecommand \@ifnum [1]{%
 \ifnum #1\expandafter \@firstoftwo
 \else \expandafter \@secondoftwo
 \fi
}%
\providecommand \@ifx [1]{%
 \ifx #1\expandafter \@firstoftwo
 \else \expandafter \@secondoftwo
 \fi
}%
\providecommand \natexlab [1]{#1}%
\providecommand \enquote  [1]{``#1''}%
\providecommand \bibnamefont  [1]{#1}%
\providecommand \bibfnamefont [1]{#1}%
\providecommand \citenamefont [1]{#1}%
\providecommand \href@noop [0]{\@secondoftwo}%
\providecommand \href [0]{\begingroup \@sanitize@url \@href}%
\providecommand \@href[1]{\@@startlink{#1}\@@href}%
\providecommand \@@href[1]{\endgroup#1\@@endlink}%
\providecommand \@sanitize@url [0]{\catcode `\\12\catcode `\$12\catcode
  `\&12\catcode `\#12\catcode `\^12\catcode `\_12\catcode `\%12\relax}%
\providecommand \@@startlink[1]{}%
\providecommand \@@endlink[0]{}%
\providecommand \url  [0]{\begingroup\@sanitize@url \@url }%
\providecommand \@url [1]{\endgroup\@href {#1}{\urlprefix }}%
\providecommand \urlprefix  [0]{URL }%
\providecommand \Eprint [0]{\href }%
\providecommand \doibase [0]{http://dx.doi.org/}%
\providecommand \selectlanguage [0]{\@gobble}%
\providecommand \bibinfo  [0]{\@secondoftwo}%
\providecommand \bibfield  [0]{\@secondoftwo}%
\providecommand \translation [1]{[#1]}%
\providecommand \BibitemOpen [0]{}%
\providecommand \bibitemStop [0]{}%
\providecommand \bibitemNoStop [0]{.\EOS\space}%
\providecommand \EOS [0]{\spacefactor3000\relax}%
\providecommand \BibitemShut  [1]{\csname bibitem#1\endcsname}%
\let\auto@bib@innerbib\@empty
\bibitem [{\citenamefont {{Seljak}}(2000)}]{2000MNRAS.318..203S}%
  \BibitemOpen
  \bibfield  {author} {\bibinfo {author} {\bibfnamefont {U.}~\bibnamefont
  {{Seljak}}},\ }\href {\doibase 10.1046/j.1365-8711.2000.03715.x} {\bibfield
  {journal} {\bibinfo  {journal} {Monthly Notices of the RAS}\ }\textbf
  {\bibinfo {volume} {318}},\ \bibinfo {pages} {203} (\bibinfo {year}
  {2000})},\ \Eprint {http://arxiv.org/abs/astro-ph/0001493}
  {arXiv:astro-ph/0001493 [astro-ph]} \BibitemShut {NoStop}%
\bibitem [{\citenamefont {Cooray}\ and\ \citenamefont
  {Sheth}(2002)}]{Cooray:2002dia}%
  \BibitemOpen
  \bibfield  {author} {\bibinfo {author} {\bibfnamefont {A.}~\bibnamefont
  {Cooray}}\ and\ \bibinfo {author} {\bibfnamefont {R.~K.}\ \bibnamefont
  {Sheth}},\ }\href {\doibase 10.1016/S0370-1573(02)00276-4} {\bibfield
  {journal} {\bibinfo  {journal} {Phys. Rept.}\ }\textbf {\bibinfo {volume}
  {372}},\ \bibinfo {pages} {1} (\bibinfo {year} {2002})},\ \Eprint
  {http://arxiv.org/abs/astro-ph/0206508} {arXiv:astro-ph/0206508 [astro-ph]}
  \BibitemShut {NoStop}%
\bibitem [{\citenamefont {{Sinha}}\ \emph {et~al.}(2018)\citenamefont
  {{Sinha}}, \citenamefont {{Berlind}}, \citenamefont {{McBride}},
  \citenamefont {{Scoccimarro}}, \citenamefont {{Piscionere}},\ and\
  \citenamefont {{Wibking}}}]{2018MNRAS.478.1042S}%
  \BibitemOpen
  \bibfield  {author} {\bibinfo {author} {\bibfnamefont {M.}~\bibnamefont
  {{Sinha}}}, \bibinfo {author} {\bibfnamefont {A.~A.}\ \bibnamefont
  {{Berlind}}}, \bibinfo {author} {\bibfnamefont {C.~K.}\ \bibnamefont
  {{McBride}}}, \bibinfo {author} {\bibfnamefont {R.}~\bibnamefont
  {{Scoccimarro}}}, \bibinfo {author} {\bibfnamefont {J.~A.}\ \bibnamefont
  {{Piscionere}}}, \ and\ \bibinfo {author} {\bibfnamefont {B.~D.}\
  \bibnamefont {{Wibking}}},\ }\href {\doibase 10.1093/mnras/sty967} {\bibfield
   {journal} {\bibinfo  {journal} {Monthly Notices of the RAS}\ }\textbf
  {\bibinfo {volume} {478}},\ \bibinfo {pages} {1042} (\bibinfo {year}
  {2018})},\ \Eprint {http://arxiv.org/abs/1708.04892} {arXiv:1708.04892
  [astro-ph.CO]} \BibitemShut {NoStop}%
\bibitem [{\citenamefont {{Ginzburg}}\ \emph {et~al.}(2017)\citenamefont
  {{Ginzburg}}, \citenamefont {{Desjacques}},\ and\ \citenamefont
  {{Chan}}}]{2017PhRvD..96h3528G}%
  \BibitemOpen
  \bibfield  {author} {\bibinfo {author} {\bibfnamefont {D.}~\bibnamefont
  {{Ginzburg}}}, \bibinfo {author} {\bibfnamefont {V.}~\bibnamefont
  {{Desjacques}}}, \ and\ \bibinfo {author} {\bibfnamefont {K.~C.}\
  \bibnamefont {{Chan}}},\ }\href {\doibase 10.1103/PhysRevD.96.083528}
  {\bibfield  {journal} {\bibinfo  {journal} {\prd}\ }\textbf {\bibinfo
  {volume} {96}},\ \bibinfo {eid} {083528} (\bibinfo {year} {2017})},\ \Eprint
  {http://arxiv.org/abs/1706.08738} {arXiv:1706.08738 [astro-ph.CO]}
  \BibitemShut {NoStop}%
\bibitem [{\citenamefont {Takahashi}\ \emph {et~al.}(2012)\citenamefont
  {Takahashi}, \citenamefont {Sato}, \citenamefont {Nishimichi}, \citenamefont
  {Taruya},\ and\ \citenamefont {Oguri}}]{Takahashi:2012em}%
  \BibitemOpen
  \bibfield  {author} {\bibinfo {author} {\bibfnamefont {R.}~\bibnamefont
  {Takahashi}}, \bibinfo {author} {\bibfnamefont {M.}~\bibnamefont {Sato}},
  \bibinfo {author} {\bibfnamefont {T.}~\bibnamefont {Nishimichi}}, \bibinfo
  {author} {\bibfnamefont {A.}~\bibnamefont {Taruya}}, \ and\ \bibinfo {author}
  {\bibfnamefont {M.}~\bibnamefont {Oguri}},\ }\href {\doibase
  10.1088/0004-637X/761/2/152} {\bibfield  {journal} {\bibinfo  {journal}
  {Astrophys. J.}\ }\textbf {\bibinfo {volume} {761}},\ \bibinfo {pages} {152}
  (\bibinfo {year} {2012})},\ \Eprint {http://arxiv.org/abs/1208.2701}
  {arXiv:1208.2701 [astro-ph.CO]} \BibitemShut {NoStop}%
\bibitem [{\citenamefont {Navarro}\ \emph {et~al.}(1996)\citenamefont
  {Navarro}, \citenamefont {Frenk},\ and\ \citenamefont
  {White}}]{Navarro:1995iw}%
  \BibitemOpen
  \bibfield  {author} {\bibinfo {author} {\bibfnamefont {J.~F.}\ \bibnamefont
  {Navarro}}, \bibinfo {author} {\bibfnamefont {C.~S.}\ \bibnamefont {Frenk}},
  \ and\ \bibinfo {author} {\bibfnamefont {S.~D.~M.}\ \bibnamefont {White}},\
  }\href {\doibase 10.1086/177173} {\bibfield  {journal} {\bibinfo  {journal}
  {Astrophys. J.}\ }\textbf {\bibinfo {volume} {462}},\ \bibinfo {pages} {563}
  (\bibinfo {year} {1996})},\ \Eprint {http://arxiv.org/abs/astro-ph/9508025}
  {arXiv:astro-ph/9508025 [astro-ph]} \BibitemShut {NoStop}%
\bibitem [{\citenamefont {Graham}\ \emph {et~al.}(2006)\citenamefont {Graham},
  \citenamefont {Merritt}, \citenamefont {Moore}, \citenamefont {Diemand},\
  and\ \citenamefont {Terzic}}]{Graham:2005xx}%
  \BibitemOpen
  \bibfield  {author} {\bibinfo {author} {\bibfnamefont {A.~W.}\ \bibnamefont
  {Graham}}, \bibinfo {author} {\bibfnamefont {D.}~\bibnamefont {Merritt}},
  \bibinfo {author} {\bibfnamefont {B.}~\bibnamefont {Moore}}, \bibinfo
  {author} {\bibfnamefont {J.}~\bibnamefont {Diemand}}, \ and\ \bibinfo
  {author} {\bibfnamefont {B.}~\bibnamefont {Terzic}},\ }\href {\doibase
  10.1086/508988} {\bibfield  {journal} {\bibinfo  {journal} {Astron. J.}\
  }\textbf {\bibinfo {volume} {132}},\ \bibinfo {pages} {2685} (\bibinfo {year}
  {2006})},\ \Eprint {http://arxiv.org/abs/astro-ph/0509417}
  {arXiv:astro-ph/0509417 [astro-ph]} \BibitemShut {NoStop}%
\bibitem [{\citenamefont {Tavio}\ \emph {et~al.}(2008)\citenamefont {Tavio},
  \citenamefont {Cuesta}, \citenamefont {Prada}, \citenamefont {Klypin},\ and\
  \citenamefont {Sanchez-Conde}}]{Tavio:2008em}%
  \BibitemOpen
  \bibfield  {author} {\bibinfo {author} {\bibfnamefont {H.}~\bibnamefont
  {Tavio}}, \bibinfo {author} {\bibfnamefont {A.~J.}\ \bibnamefont {Cuesta}},
  \bibinfo {author} {\bibfnamefont {F.}~\bibnamefont {Prada}}, \bibinfo
  {author} {\bibfnamefont {A.~A.}\ \bibnamefont {Klypin}}, \ and\ \bibinfo
  {author} {\bibfnamefont {M.~A.}\ \bibnamefont {Sanchez-Conde}},\ }\href@noop
  {} {\  (\bibinfo {year} {2008})},\ \Eprint {http://arxiv.org/abs/0807.3027}
  {arXiv:0807.3027 [astro-ph]} \BibitemShut {NoStop}%
\bibitem [{\citenamefont {Hadzhiyska}\ \emph {et~al.}(2019)\citenamefont
  {Hadzhiyska}, \citenamefont {Bose}, \citenamefont {Eisenstein}, \citenamefont
  {Hernquist},\ and\ \citenamefont {Spergel}}]{Hadzhiyska:2019xnf}%
  \BibitemOpen
  \bibfield  {author} {\bibinfo {author} {\bibfnamefont {B.}~\bibnamefont
  {Hadzhiyska}}, \bibinfo {author} {\bibfnamefont {S.}~\bibnamefont {Bose}},
  \bibinfo {author} {\bibfnamefont {D.}~\bibnamefont {Eisenstein}}, \bibinfo
  {author} {\bibfnamefont {L.}~\bibnamefont {Hernquist}}, \ and\ \bibinfo
  {author} {\bibfnamefont {D.~N.}\ \bibnamefont {Spergel}},\ }\href@noop {} {\
  (\bibinfo {year} {2019})},\ \Eprint {http://arxiv.org/abs/1911.02610}
  {arXiv:1911.02610 [astro-ph.CO]} \BibitemShut {NoStop}%
\bibitem [{\citenamefont {Okoli}\ and\ \citenamefont
  {Afshordi}(2016)}]{Okoli:2015dta}%
  \BibitemOpen
  \bibfield  {author} {\bibinfo {author} {\bibfnamefont {C.}~\bibnamefont
  {Okoli}}\ and\ \bibinfo {author} {\bibfnamefont {N.}~\bibnamefont
  {Afshordi}},\ }\href {\doibase 10.1093/mnras/stv2905} {\bibfield  {journal}
  {\bibinfo  {journal} {Mon. Not. Roy. Astron. Soc.}\ }\textbf {\bibinfo
  {volume} {456}},\ \bibinfo {pages} {3068} (\bibinfo {year} {2016})},\ \Eprint
  {http://arxiv.org/abs/1510.03868} {arXiv:1510.03868 [astro-ph.CO]}
  \BibitemShut {NoStop}%
\bibitem [{\citenamefont {{Tinker}}\ \emph {et~al.}(2010)\citenamefont
  {{Tinker}}, \citenamefont {{Robertson}}, \citenamefont {{Kravtsov}},
  \citenamefont {{Klypin}}, \citenamefont {{Warren}}, \citenamefont {{Yepes}},\
  and\ \citenamefont {{Gottl{\"o}ber}}}]{2010ApJ...724..878T}%
  \BibitemOpen
  \bibfield  {author} {\bibinfo {author} {\bibfnamefont {J.~L.}\ \bibnamefont
  {{Tinker}}}, \bibinfo {author} {\bibfnamefont {B.~E.}\ \bibnamefont
  {{Robertson}}}, \bibinfo {author} {\bibfnamefont {A.~V.}\ \bibnamefont
  {{Kravtsov}}}, \bibinfo {author} {\bibfnamefont {A.}~\bibnamefont
  {{Klypin}}}, \bibinfo {author} {\bibfnamefont {M.~S.}\ \bibnamefont
  {{Warren}}}, \bibinfo {author} {\bibfnamefont {G.}~\bibnamefont {{Yepes}}}, \
  and\ \bibinfo {author} {\bibfnamefont {S.}~\bibnamefont {{Gottl{\"o}ber}}},\
  }\href {\doibase 10.1088/0004-637X/724/2/878} {\bibfield  {journal} {\bibinfo
   {journal} {\apj}\ }\textbf {\bibinfo {volume} {724}},\ \bibinfo {pages}
  {878} (\bibinfo {year} {2010})},\ \Eprint {http://arxiv.org/abs/1001.3162}
  {arXiv:1001.3162 [astro-ph.CO]} \BibitemShut {NoStop}%
\bibitem [{\citenamefont {Tinker}\ \emph {et~al.}(2008)\citenamefont {Tinker},
  \citenamefont {Kravtsov}, \citenamefont {Klypin}, \citenamefont {Abazajian},
  \citenamefont {Warren}, \citenamefont {Yepes}, \citenamefont {Gottlober},\
  and\ \citenamefont {Holz}}]{Tinker:2008ff}%
  \BibitemOpen
  \bibfield  {author} {\bibinfo {author} {\bibfnamefont {J.~L.}\ \bibnamefont
  {Tinker}}, \bibinfo {author} {\bibfnamefont {A.~V.}\ \bibnamefont
  {Kravtsov}}, \bibinfo {author} {\bibfnamefont {A.}~\bibnamefont {Klypin}},
  \bibinfo {author} {\bibfnamefont {K.}~\bibnamefont {Abazajian}}, \bibinfo
  {author} {\bibfnamefont {M.~S.}\ \bibnamefont {Warren}}, \bibinfo {author}
  {\bibfnamefont {G.}~\bibnamefont {Yepes}}, \bibinfo {author} {\bibfnamefont
  {S.}~\bibnamefont {Gottlober}}, \ and\ \bibinfo {author} {\bibfnamefont
  {D.~E.}\ \bibnamefont {Holz}},\ }\href {\doibase 10.1086/591439} {\bibfield
  {journal} {\bibinfo  {journal} {Astrophys. J.}\ }\textbf {\bibinfo {volume}
  {688}},\ \bibinfo {pages} {709} (\bibinfo {year} {2008})},\ \Eprint
  {http://arxiv.org/abs/0803.2706} {arXiv:0803.2706 [astro-ph]} \BibitemShut
  {NoStop}%
\bibitem [{\citenamefont {Peacock}\ and\ \citenamefont
  {Smith}(2000)}]{Peacock:2000qk}%
  \BibitemOpen
  \bibfield  {author} {\bibinfo {author} {\bibfnamefont {J.~A.}\ \bibnamefont
  {Peacock}}\ and\ \bibinfo {author} {\bibfnamefont {R.~E.}\ \bibnamefont
  {Smith}},\ }\href {\doibase 10.1046/j.1365-8711.2000.03779.x} {\bibfield
  {journal} {\bibinfo  {journal} {Mon. Not. Roy. Astron. Soc.}\ }\textbf
  {\bibinfo {volume} {318}},\ \bibinfo {pages} {1144} (\bibinfo {year}
  {2000})},\ \Eprint {http://arxiv.org/abs/astro-ph/0005010}
  {arXiv:astro-ph/0005010 [astro-ph]} \BibitemShut {NoStop}%
\bibitem [{Note1()}]{Note1}%
  \BibitemOpen
  \bibinfo {note} {\protect \url {https://camb.info/}}\BibitemShut {NoStop}%
\bibitem [{\citenamefont {Lewis}\ \emph {et~al.}(2000)\citenamefont {Lewis},
  \citenamefont {Challinor},\ and\ \citenamefont {Lasenby}}]{Lewis:1999bs}%
  \BibitemOpen
  \bibfield  {author} {\bibinfo {author} {\bibfnamefont {A.}~\bibnamefont
  {Lewis}}, \bibinfo {author} {\bibfnamefont {A.}~\bibnamefont {Challinor}}, \
  and\ \bibinfo {author} {\bibfnamefont {A.}~\bibnamefont {Lasenby}},\ }\href
  {\doibase 10.1086/309179} {\bibfield  {journal} {\bibinfo  {journal} {\apj}\
  }\textbf {\bibinfo {volume} {538}},\ \bibinfo {pages} {473} (\bibinfo {year}
  {2000})},\ \Eprint {http://arxiv.org/abs/astro-ph/9911177}
  {arXiv:astro-ph/9911177 [astro-ph]} \BibitemShut {NoStop}%
\bibitem [{\citenamefont {Diemer}\ and\ \citenamefont
  {Kravtsov}(2014)}]{Diemer:2014xya}%
  \BibitemOpen
  \bibfield  {author} {\bibinfo {author} {\bibfnamefont {B.}~\bibnamefont
  {Diemer}}\ and\ \bibinfo {author} {\bibfnamefont {A.~V.}\ \bibnamefont
  {Kravtsov}},\ }\href {\doibase 10.1088/0004-637X/789/1/1} {\bibfield
  {journal} {\bibinfo  {journal} {Astrophys. J.}\ }\textbf {\bibinfo {volume}
  {789}},\ \bibinfo {pages} {1} (\bibinfo {year} {2014})},\ \Eprint
  {http://arxiv.org/abs/1401.1216} {arXiv:1401.1216 [astro-ph.CO]} \BibitemShut
  {NoStop}%
\bibitem [{\citenamefont {More}\ \emph {et~al.}(2015)\citenamefont {More},
  \citenamefont {Diemer},\ and\ \citenamefont {Kravtsov}}]{More:2015ufa}%
  \BibitemOpen
  \bibfield  {author} {\bibinfo {author} {\bibfnamefont {S.}~\bibnamefont
  {More}}, \bibinfo {author} {\bibfnamefont {B.}~\bibnamefont {Diemer}}, \ and\
  \bibinfo {author} {\bibfnamefont {A.}~\bibnamefont {Kravtsov}},\ }\href
  {\doibase 10.1088/0004-637X/810/1/36} {\bibfield  {journal} {\bibinfo
  {journal} {Astrophys. J.}\ }\textbf {\bibinfo {volume} {810}},\ \bibinfo
  {pages} {36} (\bibinfo {year} {2015})},\ \Eprint
  {http://arxiv.org/abs/1504.05591} {arXiv:1504.05591 [astro-ph.CO]}
  \BibitemShut {NoStop}%
\bibitem [{\citenamefont {Sherwin}\ \emph {et~al.}(2017)\citenamefont {Sherwin}
  \emph {et~al.}}]{Sherwin:2016tyf}%
  \BibitemOpen
  \bibfield  {author} {\bibinfo {author} {\bibfnamefont {B.~D.}\ \bibnamefont
  {Sherwin}} \emph {et~al.},\ }\href {\doibase 10.1103/PhysRevD.95.123529}
  {\bibfield  {journal} {\bibinfo  {journal} {Phys. Rev.}\ }\textbf {\bibinfo
  {volume} {D95}},\ \bibinfo {pages} {123529} (\bibinfo {year} {2017})},\
  \Eprint {http://arxiv.org/abs/1611.09753} {arXiv:1611.09753 [astro-ph.CO]}
  \BibitemShut {NoStop}%
\bibitem [{\citenamefont {Omori}\ \emph {et~al.}(2017)\citenamefont {Omori}
  \emph {et~al.}}]{Omori:2017tae}%
  \BibitemOpen
  \bibfield  {author} {\bibinfo {author} {\bibfnamefont {Y.}~\bibnamefont
  {Omori}} \emph {et~al.},\ }\href {\doibase 10.3847/1538-4357/aa8d1d}
  {\bibfield  {journal} {\bibinfo  {journal} {Astrophys. J.}\ }\textbf
  {\bibinfo {volume} {849}},\ \bibinfo {pages} {124} (\bibinfo {year}
  {2017})},\ \Eprint {http://arxiv.org/abs/1705.00743} {arXiv:1705.00743
  [astro-ph.CO]} \BibitemShut {NoStop}%
\bibitem [{\citenamefont {Aghanim}\ \emph
  {et~al.}(2018{\natexlab{a}})\citenamefont {Aghanim} \emph
  {et~al.}}]{Aghanim:2018oex}%
  \BibitemOpen
  \bibfield  {author} {\bibinfo {author} {\bibfnamefont {N.}~\bibnamefont
  {Aghanim}} \emph {et~al.} (\bibinfo {collaboration} {Planck}),\ }\href@noop
  {} {\  (\bibinfo {year} {2018}{\natexlab{a}})},\ \Eprint
  {http://arxiv.org/abs/1807.06210} {arXiv:1807.06210 [astro-ph.CO]}
  \BibitemShut {NoStop}%
\bibitem [{\citenamefont {Ade}\ \emph {et~al.}(2016)\citenamefont {Ade} \emph
  {et~al.}}]{Ade:2015zua}%
  \BibitemOpen
  \bibfield  {author} {\bibinfo {author} {\bibfnamefont {P.~A.~R.}\
  \bibnamefont {Ade}} \emph {et~al.} (\bibinfo {collaboration} {Planck}),\
  }\href {\doibase 10.1051/0004-6361/201525941} {\bibfield  {journal} {\bibinfo
   {journal} {Astron. Astrophys.}\ }\textbf {\bibinfo {volume} {594}},\
  \bibinfo {pages} {A15} (\bibinfo {year} {2016})},\ \Eprint
  {http://arxiv.org/abs/1502.01591} {arXiv:1502.01591 [astro-ph.CO]}
  \BibitemShut {NoStop}%
\bibitem [{\citenamefont {LoVerde}\ and\ \citenamefont
  {Afshordi}(2008)}]{LoVerde:2008re}%
  \BibitemOpen
  \bibfield  {author} {\bibinfo {author} {\bibfnamefont {M.}~\bibnamefont
  {LoVerde}}\ and\ \bibinfo {author} {\bibfnamefont {N.}~\bibnamefont
  {Afshordi}},\ }\href {\doibase 10.1103/PhysRevD.78.123506} {\bibfield
  {journal} {\bibinfo  {journal} {Phys. Rev.}\ }\textbf {\bibinfo {volume}
  {D78}},\ \bibinfo {pages} {123506} (\bibinfo {year} {2008})},\ \Eprint
  {http://arxiv.org/abs/0809.5112} {arXiv:0809.5112 [astro-ph]} \BibitemShut
  {NoStop}%
\bibitem [{\citenamefont {Aghanim}\ \emph
  {et~al.}(2018{\natexlab{b}})\citenamefont {Aghanim} \emph
  {et~al.}}]{Aghanim:2018eyx}%
  \BibitemOpen
  \bibfield  {author} {\bibinfo {author} {\bibfnamefont {N.}~\bibnamefont
  {Aghanim}} \emph {et~al.} (\bibinfo {collaboration} {Planck}),\ }\href@noop
  {} {\  (\bibinfo {year} {2018}{\natexlab{b}})},\ \Eprint
  {http://arxiv.org/abs/1807.06209} {arXiv:1807.06209 [astro-ph.CO]}
  \BibitemShut {NoStop}%
\bibitem [{\citenamefont {Abazajian}\ \emph {et~al.}(2016)\citenamefont
  {Abazajian} \emph {et~al.}}]{Abazajian:2016yjj}%
  \BibitemOpen
  \bibfield  {author} {\bibinfo {author} {\bibfnamefont {K.~N.}\ \bibnamefont
  {Abazajian}} \emph {et~al.} (\bibinfo {collaboration} {CMB-S4}),\ }\href@noop
  {} {\  (\bibinfo {year} {2016})},\ \Eprint {http://arxiv.org/abs/1610.02743}
  {arXiv:1610.02743 [astro-ph.CO]} \BibitemShut {NoStop}%
\bibitem [{\citenamefont {Schmidt}(2016)}]{Schmidt:2015gwz}%
  \BibitemOpen
  \bibfield  {author} {\bibinfo {author} {\bibfnamefont {F.}~\bibnamefont
  {Schmidt}},\ }\href {\doibase 10.1103/PhysRevD.93.063512} {\bibfield
  {journal} {\bibinfo  {journal} {Phys. Rev.}\ }\textbf {\bibinfo {volume}
  {D93}},\ \bibinfo {pages} {063512} (\bibinfo {year} {2016})},\ \Eprint
  {http://arxiv.org/abs/1511.02231} {arXiv:1511.02231 [astro-ph.CO]}
  \BibitemShut {NoStop}%
\bibitem [{\citenamefont {Seljak}\ and\ \citenamefont
  {Vlah}(2015)}]{Seljak:2015rea}%
  \BibitemOpen
  \bibfield  {author} {\bibinfo {author} {\bibfnamefont {U.}~\bibnamefont
  {Seljak}}\ and\ \bibinfo {author} {\bibfnamefont {Z.}~\bibnamefont {Vlah}},\
  }\href {\doibase 10.1103/PhysRevD.91.123516} {\bibfield  {journal} {\bibinfo
  {journal} {Phys. Rev.}\ }\textbf {\bibinfo {volume} {D91}},\ \bibinfo {pages}
  {123516} (\bibinfo {year} {2015})},\ \Eprint
  {http://arxiv.org/abs/1501.07512} {arXiv:1501.07512 [astro-ph.CO]}
  \BibitemShut {NoStop}%
\bibitem [{\citenamefont {Hand}\ \emph {et~al.}(2017)\citenamefont {Hand},
  \citenamefont {Seljak}, \citenamefont {Beutler},\ and\ \citenamefont
  {Vlah}}]{Hand:2017ilm}%
  \BibitemOpen
  \bibfield  {author} {\bibinfo {author} {\bibfnamefont {N.}~\bibnamefont
  {Hand}}, \bibinfo {author} {\bibfnamefont {U.}~\bibnamefont {Seljak}},
  \bibinfo {author} {\bibfnamefont {F.}~\bibnamefont {Beutler}}, \ and\
  \bibinfo {author} {\bibfnamefont {Z.}~\bibnamefont {Vlah}},\ }\href {\doibase
  10.1088/1475-7516/2017/10/009} {\bibfield  {journal} {\bibinfo  {journal}
  {JCAP}\ }\textbf {\bibinfo {volume} {1710}},\ \bibinfo {pages} {009}
  (\bibinfo {year} {2017})},\ \Eprint {http://arxiv.org/abs/1706.02362}
  {arXiv:1706.02362 [astro-ph.CO]} \BibitemShut {NoStop}%
\bibitem [{\citenamefont {Thiele}\ \emph {et~al.}(2019)\citenamefont {Thiele},
  \citenamefont {Hill},\ and\ \citenamefont {Smith}}]{Thiele:2018jdl}%
  \BibitemOpen
  \bibfield  {author} {\bibinfo {author} {\bibfnamefont {L.}~\bibnamefont
  {Thiele}}, \bibinfo {author} {\bibfnamefont {J.~C.}\ \bibnamefont {Hill}}, \
  and\ \bibinfo {author} {\bibfnamefont {K.~M.}\ \bibnamefont {Smith}},\ }\href
  {\doibase 10.1103/PhysRevD.99.103511} {\bibfield  {journal} {\bibinfo
  {journal} {Phys. Rev.}\ }\textbf {\bibinfo {volume} {D99}},\ \bibinfo {pages}
  {103511} (\bibinfo {year} {2019})},\ \Eprint
  {http://arxiv.org/abs/1812.05584} {arXiv:1812.05584 [astro-ph.CO]}
  \BibitemShut {NoStop}%
\end{thebibliography}%

\end{document}